# Nonlocal Kinetic Energy Density Functionals for Isolated Systems via Local Density Approximation Kernels

Qiang Xu[†], Jian Lv[†], Yanchao Wang[*,†], and Yanming Ma[*,†,#]

[†] *Innovation Center of Computational Physics Methods and Software & State Key Lab of Superhard Materials, College of Physics, Jilin University, Changchun 130012, China.*

[#]*International Center of Future Science, Jilin University, Changchun 130012, China*

*Corresponding Authors:*

[*]*E-mail: wyc@calypso.cn*

[*]*E-mail: mym@calypso.cn*

Despite a large number of nonlocal kinetic energy density functionals (KEDFs) available for large-scale calculations, most of those nonlocal KEDFs designed for the extended systems cannot be directly applied to isolated systems. In this manuscript, we proposed a generalized scheme to construct nonlocal KEDFs via the local density approximation kernels and construct a family of KEDFs for simulations of isolated systems within orbital-free density functional theory. The performance of KEDFs has been demonstrated by several clusters encompassing Mg, Si and GaAs. The results show that our constructed KEDFs can achieve high numerical accuracy and stability for random clusters, therefore, making orbital-free density functional theory accessible for practical simulations of isolated systems.

# 1. Introduction

*Ab initio* calculation based on density functional theory (DFT)[1,2] as a prevalent tool for materials simulation has provided important insights into a variety of materials. Particularly, orbital-free (OF) DFT has been recognized as a practical means for large-scale simulations, as exemplified by the calculations of simple metals containing millions of atoms in simulated cell[3–5]. However, the accuracy of OF-DFT heavily depends on the approximation of kinetic energy density functional (KEDF) since the kinetic energy is the same order of magnitude as the total energy. Therefore, the main barrier to widespread use of OF-DFT is the lack of reliable KEDFs with high transferability and numerical stability.

In the past few decades, a large number of KEDFs including local/semilocal and nonlocal KEDFs have been available. The local/semilocal KEDFs such as Thomas-Fermi (TF)[6–8], von Weizsäcker (vW)[9], generalized gradient approximation[10–19], and meta-generalized gradient approximation[20,21] functionals are constructed using the local electron density or its gradient and Laplacian. These functionals can be easily applied to isolated systems[22–25]. However, local/semilocal functionals cannot reproduce the quantum oscillation of electron density, such as atomic shell structure[20,26] and Friedel oscillations[27,28]. In order to capture the quantum oscillation of electron density, several nonlocal KEDFs such as Wang-Teter (WT)[29], Smargiassi-Madden (SM)[30], Perrot[31] and Mi-Genova-Pavanello (MGP)[32], etc.[28,33] have been proposed by employment of density-independent kernels with a constant Fermi wave vector (FWV) of $k_F^0 = \left(3\pi^2 \rho_0\right)^{1/3}$. However, the constant FWV is usually related to the average density ($\rho_0$) in the unit cell for extended systems and is not well defined in isolated systems[21].

To avoid using the constant FWV, the density-dependent weight function or kernel are employed in several nonlocal KEDFs including Chacón-Alvarellos-Tarazona (CAT)[27], Wang-Govind-Carter (WGC)[34] and Huang-Carter (HC)[35]. However, they suffer from poor transferability or numerical instability problems for isolated systems[36]. Furthermore, the solution of differential equations is required to make these KEDFs recover the linear response of uniform electron gas, which is inappropriate for modeling of isolated systems. A nonlocal functional with propagator-like kernel proposed by Wang and Teter successfully reproduced the atomic shell structures[29]. However, this KEDF required artificial introduction of Gaussian functions with fitting parameters. Later, the advanced nonlocal KEDFs have been proposed and used to simulate the isolated systems[37–42]. Particularly, a family of nonlocal KEDFs named *LX* (*X*=WT, MGP0, MGP) were recently constructed using the numerical local density approximated approach. The *LX* KEDFs were proved to achieve close to chemical accuracy and high transferability for clusters[36]. Unfortunately, these KEDFs still suffer from the numerical instability in some cases.

In this manuscript, a generalized scheme has been proposed to construct KEDFs for isolated systems by introduction of the local density-dependent kernels and a variety of nonlocal KEDFs have been constructed within the scheme. We have implemented these KEDFs into ATLAS[43] for numerical calculations of isolated systems within OF-DFT. The high accuracy and numerical stability of these KEDFs have been demonstrated by successful applications to several clusters.

The remainder of this manuscript is organized as follows. Section 2 briefly gives the OF-DFT, followed by the detailed scheme for construction of KEDFs and their implementation into ATLAS. The computational details are provided in Section 3. The accuracy and numerical stability of the proposed KEDFs for simulations of isolated systems have been demonstrated in section 4. Finally, we give conclusions in Section 5.

## 2. Theory and Implementation
### 2.1 Orbital-free density functional theory

In OF-DFT, the ground-state energy $E_{GS}$ and electron density $\rho_{GS}$ are obtained by minimizing the total energy functional $E[\rho]$ of the $N_e$-electron system[44]

$$E_{GS}[\rho_{GS}] = \min_{\rho} \left\{ E[\rho] - \mu \left( \int_{\Omega} \rho(\vec{r}) d^3r - N_e \right); \rho \geq 0 \right\}, \qquad (1)$$

where $\rho$ is the electron density and $\mu$ denoting the Lagrange multiplier is used to enforce the constraint that the total number of electrons. The total energy density functional $E[\rho]$ can be written as

$$E[\rho] = T_s[\rho] + E_H[\rho] + E_{ie}[\rho] + E_{xc}[\rho] + E_{ii}(R), \qquad (2)$$

where $T_s$, $E_H$, $E_{ie}$, $E_{xc}$, $E_{ii}$ and $R$ denote terms of noninteracting kinetic energy, the Hartree energy, the ion-electron interaction energy, the exchange-correlation energy, the ion-ion repulsion energy and the collection of ionic positions, respectively. In contrast to Kohn-Sham (KS) DFT, where the exact noninteracting kinetic energy term is evaluated by single-particle orbitals, OF-DFT relies upon explicit functionals of the electron density for all energy terms.

**2.2 The nonlocal KEDFs for isolated systems**

Most of nonlocal KEDFs can be written in the generic form

$$T_s[\rho] = T_{TF}[\rho] + T_{vW}[\rho] + T_{NL}^{X}[\rho], \qquad (3)$$

where $T_{TF}[\rho] = \frac{3}{10}(3\pi^2)^{2/3}\langle \rho^{5/3}(\vec{r})\rangle$ and $T_{vW}[\rho] = \frac{1}{8}\left\langle \frac{|\nabla\rho(\vec{r})|^2}{\rho(\vec{r})}\right\rangle$ are the Thomas-Fermi[6–8] and von Weizsäcker[9] KEDFs, respectively. The last term in Eq. (3) is the nonlocal part of KEDFs. A simplest form of nonlocal part of KEDFs is expressed as Eq. (4) and includes a density-independent kernel $w_{\alpha,\beta}^{X}$.

$$T_{NL}^{X}[\rho] = \langle \rho^{\alpha}(\vec{r})|w_{\alpha,\beta}^{X}[k_F^0, \vec{r}-\vec{r}']|\rho^{\beta}(\vec{r}')\rangle, \qquad (4)$$

where $\alpha$ and $\beta$ are positive parameters that define $X$=WT[29], MGP and MGP0[32] for $\alpha = \beta = 5/6$, $X$=SM[30] for $\alpha = \beta = 1/2$ and $X$=Perrot[31] for $\alpha = \beta = 1$.

In our scheme, we reformulate the nonlocal term of KEDFs by introduction of the local density approximation kernels (LDAK). Specifically, the constant $k_F^0$ in density-independent kernel of KEDFs of Eq. (4) is directly substituted by local FWV of $k_F(\vec{r}) = (3\pi^2\rho(\vec{r}))^{1/3}$. In other words, a density-dependent kernel related to the local electron density instead of average electron density is employed in our scheme. Within this scheme, the nonlocal terms of KEDFs in Eq. (4) are reformulated as

$$T_{NL}^{LDAK-X}[\rho]=\langle\rho^\alpha(\vec{r})|w_{\alpha,\beta}^X[k_F(\vec{r}),\vec{r}-\vec{r}']|\rho^\beta(\vec{r}')\rangle, \tag{5}$$

The corresponding kinetic energy potentials (KEPs) are given by

$$\begin{aligned}V_{T,NL}^{LDAK-X}[\rho]=&\alpha\rho^{\alpha-1}(\vec{r})\int w_{\alpha,\beta}^X[k_F(\vec{r}),\vec{r}-\vec{r}']\rho^\beta(\vec{r}')d^3r'\\&+\rho^\alpha(\vec{r})\int\frac{dw_{\alpha,\beta}^X[k_F(\vec{r}),\vec{r}-\vec{r}']}{d\rho(\vec{r})}\rho^\beta(\vec{r}')d^3r'\\&+\beta\rho^{\beta-1}(\vec{r})\int w_{\alpha,\beta}^X[k_F(\vec{r}'),\vec{r}-\vec{r}']\rho^\alpha(\vec{r}')d^3r'.\end{aligned} \tag{6}$$

**2.3 The implementation of OF-DFT for isolated systems**

The previous version of ATLAS code has been used for numerical calculations of periodic systems within OF-DFT[43]. The long-range electrostatic interactions (ion-ion, ion-electron, and electron-electron interactions) under the periodic boundary conditions are evaluated by introduction of an artificial supercell with large vacuum for isolated systems[45–47]. However, it usually leads to slow convergence of the total energy with supercell size if there are the strong multipole-multipole interactions between the periodic replicas[48].

Herein, a capability for simulations of isolated clusters has been implemented in ATLAS code, where the long-range electrostatic interactions are calculated under the Dirichlet boundary condition (DBC). In general, all the electrostatic energy terms can be calculated with a linear scaling under DBC except for the ion-ion interaction term. The ion-ion interaction energy in Eq. (2) is defined as

$$E_{ii}(R)=\sum_{I=1}^{N_a}\sum_{J>I}^{N_a}\frac{Z_IZ_J}{R_{IJ}}, \tag{7}$$

where $N_a$ is the number of atoms, $R_{IJ}=|\vec{R}_I-\vec{R}_J|$. $\{\vec{R}_I\}$ and $\{Z_I\}$ denote the ionic positions and charges, respectively. Obviously, a direct calculation of ion-ion interaction shows an intrinsic square scaling with respect to the number of ions. In fact, the Eq. (2) can also be reformulated as[49,50]

$$E[\rho] = T_s[\rho] + E_{xc}[\rho] + E_{ele}[\rho, R], \tag{8}$$

where $E_{ele}$ denoting the electrostatic interaction energy contains the ion-ion, ion-electron, and electron-electron interactions. The electrostatics can be expressed by[49,51,52]:

$$E_{ele}[\rho, R] = \sup_{V_{ele}} \left\{ -\frac{1}{8\pi} \int |\nabla V_{ele}(\vec{r})|^2 d^3r + \int (\rho(\vec{r}) + b(\vec{r})) V_{ele}(\vec{r}) d^3r \right\}$$
$$- E_{self}(R) + E_c(R), \tag{9}$$

where $V_{ele}$ is referred as the electrostatic potential, $b$ is the total pseudo-charge density of the nuclei, $E_{self}$ is the self-energy of nuclei, $E_c$ is used to correct the error of ion-ion repulsive energy due to overlap of pseudo-charge density. The electrostatic potential $V_{ele}$ in Eq. (9) is calculated by solving the Poisson equation:

$$\nabla^2 V_{ele}[n](\vec{r}) = -4\pi n(\vec{r}). \tag{10}$$

The total density of $n$ is defined as the sum of pseudo-charge density and electron density

$$n(\vec{r}) = \rho(\vec{r}) + b(\vec{r}). \tag{11}$$

The detailed calculations of $b$, $E_{self}$ and $E_c$ can be found in Ref. 49.

In this work, the electron density distribution and the corresponding electrostatic potentials are represented on real-space discrete Cartesian grid points. Just as shown in Fig. 1(a), the radius $R_{max}$ of a spherical region is used to truncate the tail of electron density, whose value should be zero beyond the spherical region. The electrostatic potentials are directly represented on discretized grid points in cubic cell. Note that the unit cell length edge is defined as $L = 2R_{max}$. There are two types of grid points in our implementation as illustrated in Fig. 1(b). The electrostatic potentials on boundary points are calculated by the multipoles expansion method[53], whereas the electrostatic potentials on the internal points can be solved by conjugate gradient iteration with multigrid in real space[54]. The number of boundary layers is determined by the order of finite difference. For efficient linear-scaling calculations of $T_{NL}^{LDAK-X}[\rho]$ and the corresponding potentials $V_{T,NL}^{LDAK-X}[\rho]$, the integrals of $P(\vec{r})=\int w[k_F(\vec{r}),|\vec{r}-\vec{r}'|]f(\vec{r}')d^3r'$ and $Q(\vec{r})=\int w[k_F(\vec{r}'),|\vec{r}-\vec{r}'|]f(\vec{r}')d^3r'$ in Eqs. (5) and (6) are calculated by cubic Hermite spline interpolation technique and fast Fourier transform (FFT)[35]. It is important to note that the computational cost of Eqs. (5) and (6) becomes intrinsic quasilinear scaling $O[mN\log N]$. Note that $m$ and $N$ are the number of uniform interpolation nodes of FVWs and FFT grids, respectively. The details of these techniques are provided in Ref. 35. The ground-state electron density is obtained by minimizing the total energy using the truncated Newton method[55] and more details can be found in Ref. 43.

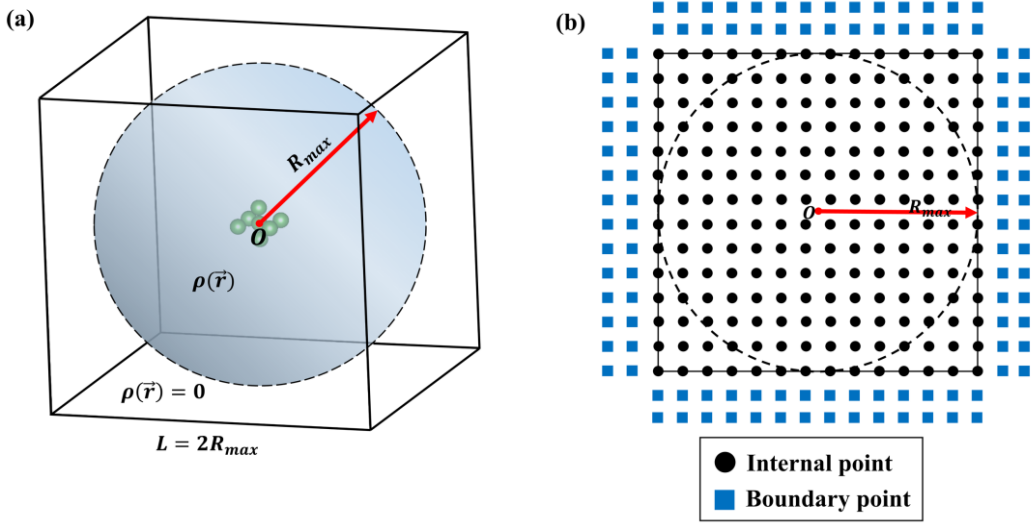

Fig. 1. The schematic illustrations of ATLAS implementation. (a) The electron density distribution in cubic cell. (b) Two-dimensional diagram showing the different types of discretized grid points.

## 3. Computational details

The OF-DFT calculations with LDAK-X and LX functionals were carried out by ATLAS. A grid spacing of 0.2 Å and eighth finite-difference order gave well convergence of total energies less than 1 meV/atom. The parameter $A=0.2$ of MGP[36] was kept fixed for both LDAK-MGP and LMGP. The number of interpolation nodes of 40 and 100 in LDAK-$X$ for clusters of Mg and Si/GaAs gave total energies convergence within 5 meV/atom. Calculations involving the CAT KEDF, in which kinetic energy cutoff of 1600eV, $\rho^* = 0.20\,\text{Å}^{-3}$ and $\gamma = 1.4$, are performed with PROFESS 3.0[3]. The KS-DFT calculations were performed by in-house developed ARES software package[56] and double checked using CASTEP[57]. A grid spacing of 0.2 Å and 16th finite-difference order in ARES and kinetic energy cutoff of 940 eV for CASTEP were sufficient for a well-converged total energy (1 meV/atom). The bulk-derived local pseudopotentials[58] and local density approximate exchange and correlation as parametrized by Perdew and Zunger[59] were employed to estimate the ion-electron and the exchange-correlation interactions for all the considered systems. The structures of $Mg_8$, $Mg_{50}$, $Ga_4As_4$, $Ga_{25}As_{25}$ and $Si_{50}$ were randomly generated by CALYPSO[60,61]. The settings of $R_{max} = 9.5$ Å for $Mg_8$ and $Ga_4As_4$ and $R_{max} = 13.0$ Å for $Mg_{50}$, $Ga_{25}As_{25}$, $Si_{50}$ and $Si_{60}$ yielded good convergence of total energy.

## 4. Results and Discussion

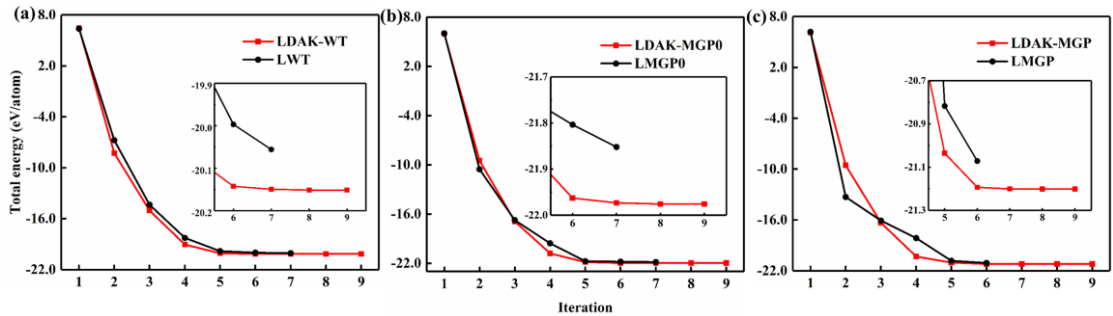

Fig. 2. The comparison of total energy convergence for $Mg_8$ between LDAK-$X$ and L$X$, where $X$ denotes (a) WT, (b) MGP0 and (c) MGP, respectively.

To assess the performance of our LDAK-$X$ scheme, we firstly construct a family of KEDFs and perform the energy minimization of $Mg_8$ using OF-DFT with these KEDFs. For comparison, we also include the results of L$X$ ($X$=WT, MGP0 and MGP).

Just as shown in Fig. 2, our LDAK-*X* KEDFs show a significant improvement of numerical stability in comparison with L*X* KEDFs. For example, it only requires seven iterations to give total energy convergence less than 1 meV/atom for a random structure of Mg$_8$ using LDAK-MGP in Fig. 2(c), whereas it fails to converge using LMGP[36].

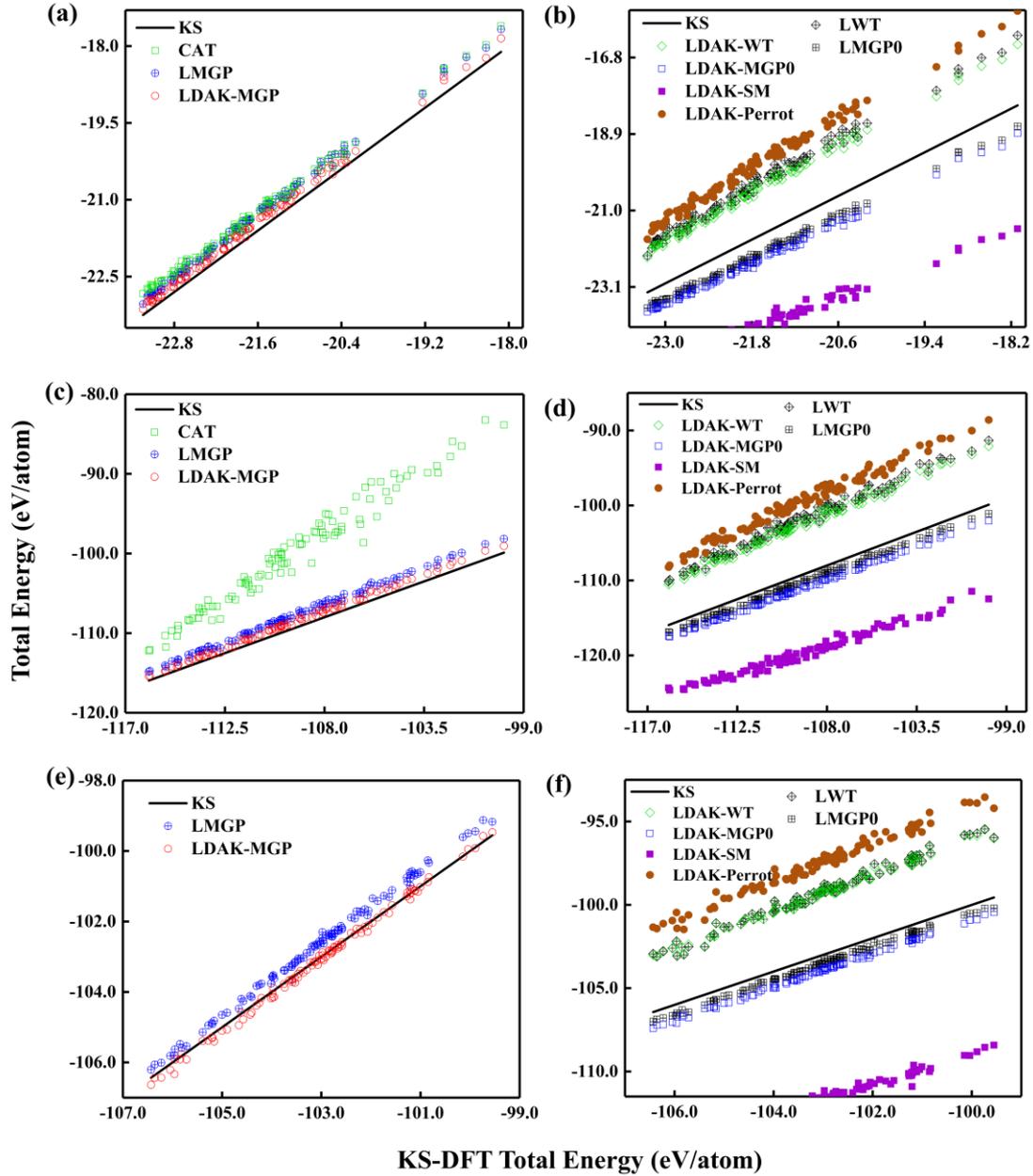

Fig. 3. The total energies of 100 random clusters calculated by OF-DFT with a variety of KEDFs in comparison with the reference KS-DFT results for (a)-(b) Mg$_8$, (c)-(d) Ga$_4$As$_4$ and (e)-(f) Si$_{50}$, respectively.

The different converged behaviors of LDAK-*X* and L*X* KEDFs originate from their different mathematic frameworks. Just as presented in Eqs. (6) and (12), the formulas of KEPs for LDAK-$X$ and L$X$ schemes are remarkably different. In the L$X$ scheme, the KEPs are calculated by spline interpolation[36]

$$V_{T,NL}^{LX}[\rho](\vec{r}) = \frac{5}{3}\rho^{-1/6}(\vec{r})\int w^X\left[k_F(\vec{r}),|\vec{r}-\vec{r}'|\right]\rho^{5/6}(\vec{r}')d^3r'$$

$$= \frac{5}{3}\rho^{-1/6}(\vec{r})\sum_{i=1}^{m}c_i\left[\rho(\vec{r}),\vec{r}\right]\int w^X\left[k_i,|\vec{r}-\vec{r}'|\right]\rho^{5/6}(\vec{r}')d^3r', \qquad (12)$$

where $\{c_i\}$ denotes the spline interpolation coefficients and $\{k_i\}$ is the set of interpolation nodes. Note that those coefficients depend on the local density $\rho(\vec{r})$. In general, KEDF can be obtained by direct integration of KEP. However, it suffers from high computational costs due to involving the extremely complicated integrations. A simple approximation, which regards $\{c_i\}$ as density-independent parameters, was employed in the L$X$ scheme[36] to obtain the KEDF by line integral from the KEP in Eq. (12). However, this approximation is so strong that the derivative relation between KEDF and KEP cannot be strictly satisfied. Hence L$X$ KEDFs suffer from numerical instabilities during energy minimization for some cases. In contrast, LDAK-$X$ KEDFs are constructed by direct introduction of local density dependent kernels and the corresponding KEP is obtained by derivative of KEDF. Therefore, the derivative relation between KEDF and KEP is strict, making LDAK-$X$ functionals numerically stable during energy minimization.

To evaluate the accuracy of LDAK-$X$, total energies of 100 random structures of $Mg_8$, $Ga_4As_4$ and $Si_{50}$ clusters were evaluated by OF-DFT with various KEDFs including LDAK-WT, LDAK-MGP0, LDAK-MGP, LDAK-SM, LDAK-Perrot, LWT, LMGP0, LMGP and CAT functionals. The calculated OF-DFT energies in comparison with that of KS-DFT are shown in Fig. 3. OF-DFT calculations within LDAK-*X* and L*X* KEDFs generally produce similar trends of total energies as KS-DFT for all considered systems. Especially, LDAK-MGP and LMGP show a significant improvement in computational accuracy compared to other functionals. The performance of CAT functional is quite modest for $Mg_8$ clusters [Fig. 3(a)], while the total energies obtained by CAT functional and KS-DFT show an apparent discrepancy for $Ga_4As_4$ clusters [Fig. 3(c)]. Particularly, the energy minimization of random structures of $Si_{50}$ clusters fails to converge using CAT functional.

Table 1. The mean-unsigned-error (MUE) of the total energies (eV/atom) and the mean-unsigned-relative-error (MURE) of electron density in percentage points (in parentheses) with respect to the KS-DFT results for 100 random structures of $Mg_8$, $Ga_4As_4$ and $Si_{50}$. The underline highlights the results close to KS-DFT.

| KEDF | MUE of energy (MURE of density) | | |
| --- | --- | --- | --- |
| | $Mg_8$ | $Ga_4As_4$ | $Si_{50}$ |
| LWT | 1.444 ( 7.8) | 7.281 ( 8.1) | 3.744 ( 4.4) |
| LMGP0 | 0.501 ( 8.8) | 1.054 (10.3) | 0.512 ( 8.3) |
| LMGP | 0.313 ( 7.7) | 1.528 (10.0) | 0.457 ( 8.6) |
| LDAK-SM | 2.406 (31.2) | 10.444 (30.6) | 8.547 (29.8) |
| LDAK-Perrot | 1.969 ( 9.9) | 9.649 (12.7) | 5.535 ( 8.9) |
| LDAK-WT | 1.295 ( 8.3) | 6.719 ( 9.9) | 3.701 ( 7.6) |
| LDAK-MGP0 | 0.650 ( 7.1) | 1.818 ( 6.5) | 0.931 ( <u>3.5</u>) |
| LDAK-MGP | <u>0.164</u> ( <u>6.9</u>) | <u>0.766</u> ( <u>6.3</u>) | <u>0.086</u> ( <u>3.5</u>) |
| CAT | 0.370 ( 9.7) | 9.522 (17.5) | - |

In order to further quantify the accuracy of KEDFs, we defined unsigned-error of total energy $\Delta E_i = \frac{1}{N_a}\left|E_i^{OF} - E_i^{KS}\right|$ and unsigned-relative-error of electron density $\Delta \rho_i = \frac{1}{N_e}\int\left|\rho_i^{OF}(\vec{r}) - \rho_i^{KS}(\vec{r})\right|d^3r$ for *i-th* cluster. The mean-unsigned-error of total energies $\overline{\Delta E} = \frac{1}{100}\sum_{i=1}^{100}\Delta E_i$ and the mean-unsigned-relative-error of electron density

$$\overline{\Delta\rho} = \frac{1}{100}\sum_{i=1}^{100}\Delta\rho_i$$ for 100 random structures of $Mg_8$, $Ga_4As_4$ and $Si_{50}$ are listed in Table 1. It is apparent that LDAK-MGP outperforms other KEDFs and yields the smallest $\overline{\Delta E}$ and $\overline{\Delta\rho}$ in all considered cases. Furthermore, it should be stressed that energy minimization using LDAK-MGP is able to obtain high convergence rates approaching 100% for all the random structures, which is superior to that of L$X$[36].

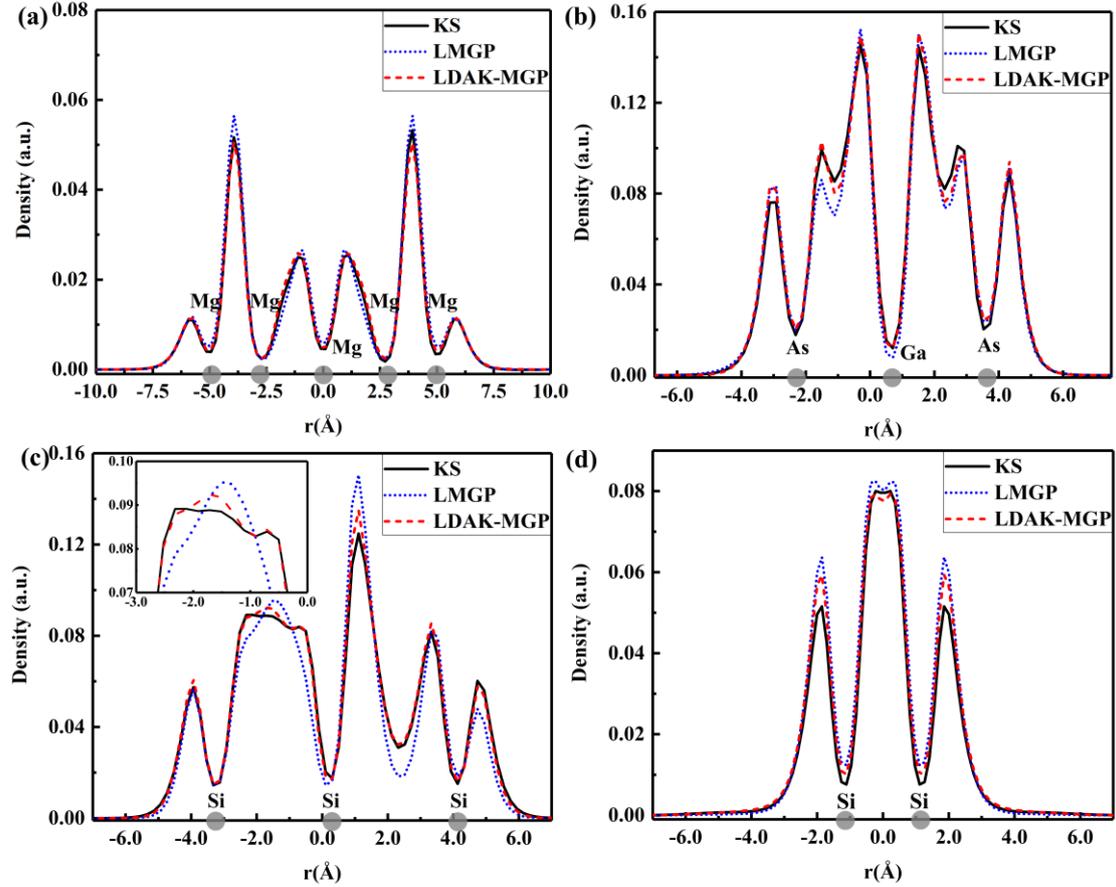

Fig. 4. The electron densities calculated by LMGP (blue dot), LDAK-MGP (red short dash) and KS-DFT (black solid line) for (a) $Mg_{50}$ (b) $Ga_{25}As_{25}$ (c) $Si_{50}$ and (d) $Si_{60}$ along the specific bond orientation.

In addition, we also evaluated electron densities of $Mg_{50}$, $Ga_{25}As_{25}$, $Si_{50}$ and $Si_{60}$ using LDAK-MGP in comparison with those estimated by LMGP, as well as KS-DFT. The detailed structural information and the corresponding directions for each structure are presented in the Supplemental Material[62]. As shown in Fig. 4, the electron density distributions predicted by LDAK-MGP shares the similar general shapes with KS-DFT in the all regions, while LMGP gives quit different distributions for the bonding regions and near-core regions. It is important to note that LDAK-MGP successfully reproduces the tiny density oscillation obtained by KS-DFT in the bonding region for $Si_{50}$, as evidenced by insert of Fig. 4(c). These results reveal that LDAK-MGP gives more accurate distributions of electron density for isolated systems than those obtained by LMGP.

## 5. Conclusion

In summary, a LDAK-$X$ scheme derived from the local density approximation is proposed to construct a family of nonlocal KEDFs for isolated systems. These KEDFs have been implemented into the ATLAS package and showed superior performance to other KEDFs both numerical accuracy and stability for several clusters encompassing Mg, Si and GaAs. The LDAK-MGP with high accuracy and numerical stability makes OF-DFT as the most promising approach for simulations of isolated systems.

**Acknowledgements** Y. W. and Y. M. acknowledge funding support from the National Key Research and Development Program of China under Grant No. 2016YFB0201201, and 2017YFB0701503; the National Natural Science Foundation of China under Grants No. 11404128, 11822404, 11534003 and 11774127; supported by Program for JLU Science and Technology Innovative Research Team (JLUSTIRT); and the Science Challenge Project, No. TZ2016001. Part of the calculation was performed in the high-performance computing center of Jilin University. We thank Mi Wenhui for helpful discussions.